# Visualizing the atomic scale electronic structure of the $Ca_2CuO_2Cl_2$ Mott insulator

Cun Ye[1,*], Peng Cai[1,*], Runze Yu[2], Xiaodong Zhou[1], Wei Ruan[1], Qingqing Liu[2], Changqing Jin[2] & Yayu Wang[1,†]

[1]*State Key Laboratory of Low Dimensional Quantum Physics, Department of Physics, Tsinghua University, Beijing 100084, China*

[2] *Institute of Physics, Chinese Academy of Sciences, Beijing 100190, China*

[*]*These authors contributed equally to this work.*

[†] E-mail: yayuwang@tsinghua.edu.cn

**Although the mechanism of superconductivity in the cuprates remains elusive, it is generally agreed that at the heart of the problem is the physics of doped Mott insulators[1]. The cuprate parent compound has one unpaired electron per Cu site, and is predicted by band theory to be a half-filled metal. The strong onsite Coulomb repulsion, however, prohibits electron hopping between neighboring sites and leads to a Mott insulator ground state with antiferromagnetic (AF) ordering[2]. Charge carriers doped into the $CuO_2$ plane destroy the insulating phase and superconductivity emerges as the carrier density is sufficiently high. The natural starting point for tackling high $T_C$ superconductivity is to elucidate the electronic structure of the parent Mott insulator and the behavior of a single doped charge. Here we use a scanning tunneling microscope to investigate the atomic scale electronic structure of the $Ca_2CuO_2Cl_2$ parent Mott insulator of the cuprates. The full electronic spectrum across the Mott-Hubbard gap is uncovered for the first time, which reveals the particle-hole symmetric and spatially uniform Hubbard bands. A single electron donated by surface defect is found to create a broad in-gap electronic state that is strongly localized in space with spatial characteristics intimately related to the AF spin background. The unprecedented real space electronic structure of the parent cuprate sheds important new light on the origin of high $T_C$ superconductivity from the doped Mott insulator perspective.**

Despite extensive experimental studies on parent and lightly doped Mott insulators, a comprehensive picture about their electronic structure is still lacking. Optical spectroscopy[3] and resonant inelastic x-ray spectroscopy (RIXS)[4-6] can probe the excitations across the Mott-Hubbard gap, but they cannot detect the ground state electronic structure directly.

Angle-resolved photoemission spectroscopy (ARPES) provides direct momentum ($k$) space structure of the occupied lower Hubbard band (LHB)[7-9], but is unable to reach the empty states above the Fermi level ($E_F$), leaving the upper Hubbard band (UHB) unexplored. The lack of complete spectroscopic information across the Mott-Hubbard gap makes it difficult to clarify how the high energy spectral weight is transferred to the low energy states near $E_F$, upon which superconductivity emerges.

Another vital piece of information that is still missing is the spatial characteristic of the Mott insulator electronic structure. Because the Mott physics is dictated by onsite Coulomb repulsion, the real space picture could be more relevant than the $k$-space picture, which has been extremely successful in describing weakly-interacting electron systems. Especially, atomic scale resolution becomes indispensible if we want to experimentally illustrate the electronic properties of a single charge doped into the AF background.

Scanning tunneling microscopy (STM) is an ideal spectroscopic technique for filling the aforementioned gaps in our understanding of doped Mott insulators. STM can directly map out the electronic states both below and above $E_F$, with the unique capability of atomic scale spatial resolution. STM studies on high $T_C$ cuprates have revealed a variety of remarkable phenomena including the impurity resonance[10,11], checkerboard-like ordering[12-15], phase incoherent pairing[16,17], and quasiparticle interference[18]. However, the parent Mott insulator phase has never been investigated by STM, most likely due to the technical challenges associated with tunneling into a good insulator. Our recent advance in STM measurements at relatively high temperature and over large bias range enables the atomic scale studies of the parent Mott insulator, as described in the Method Summary.

The oxychloride cuprate $Ca_2CuO_2Cl_2$ (CCOC) parent Mott insulator is chosen for its high crystal quality and excellent surface property. Figure 1a shows the schematic crystal structure of CCOC, which is similar to that of the prototypical parent compound $La_2CuO_4$. The crystal cleaves easily between two adjacent CaCl layers, exposing the Cl terminated surface. The $CuO_2$ plane lies at 2.73Å below the surface and the Cu sites are directly underneath the exposed apical Cl atoms. Figure 1b displays a typical STM topography of cleaved CCOC taken at $T = 77K$ with sample bias $V_s = -3.5V$ and tunneling current $I_t = 10pA$, which reveals a large area of atomically flat terrace. The Cl atoms form a regular square lattice with lattice constant $a \sim 3.9$Å, in agreement with the tetragonal unit cell of CCOC. The cross-shaped dark spots in the image correspond to the missing Cl defects, which are created during the cleaving process. The inset of Fig. 2a displays the high resolution STM image of a defect-free area of CCOC, which has arguably the most ideal surface structure for the cuprates without any complication from structural supermodulations and electronic inhomogeneities.

The local electronic structure of CCOC is probed by $dI/dV$ (differential conductance) spectroscopy. Figure 2a displays the spatially averaged $dI/dV$ curve obtained on the surface shown in the inset, which uncovers the full electronic spectrum across the Mott-Hubbard gap. The occupied state $dI/dV$ curve is very similar to the ARPES measured energy distribution curve of CCOC[8,19]. Since the parent cuprate is a charge transfer insulator with oxygen bands lying between the lower and upper Hubbard bands[20], the shoulder near $V_s = -0.8V$ is ascribed to the charge transfer band (CTB) and the rapid increase for $V_s < -2.0V$ is due to the nonbonding O $2p_\pi$ state[8,19]. The $dI/dV$ of the CTB can be fit well by a Gaussian lineshape (Fig. S2a) with the full-width at half-maximum (FWHM) $\Gamma = 0.47V$, which is in rough agreement with the bandwidth $W = 0.35$eV determined by the ARPES band dispersion[19]. The totally new feature revealed by Fig. 2a is the UHB above $E_F$, which was an unexplored territory before this report. The UHB also shows a Gaussian-like profile that peaks around $V_s = 2.3$V with $\Gamma = 0.34$eV (Fig. S2b). The simultaneous observation of CTB and UHB immediately reveals several key spectroscopic features of the Mott insulator phase. The effective Mott-Hubbard gap, or more precisely the charge transfer gap, can be directly measured to be $\Delta = 2.2$eV. This value is in good agreement with that obtained by optical spectroscopy and RIXS on parent CCOC[5]. The $E_F$ lies near the CTB edge, as has been seen in the ARPES results[19], due to the hole-type carriers in the bulk induced by defects. The $dI/dV$ lineshape and peak value of the CTB and UHB are very similar to each other, indicating approximate particle-hole symmetry across the charge transfer gap. The electronic structure of the Mott insulator phase is schematically illustrated in Fig. 2b.

We next turn to the spatial distribution of the electronic structure. Since the lowest energy excitation of the charge transfer insulator is the removal of an electron from the occupied O 2$p$ orbital to the unoccupied Cu $3d_{x^2-y^2}$ orbital[20], naïvely it is expected that the unpaired hole will mainly reside on the anion ligands and electron on the cation $d$ orbitals[21]. This picture implies that the relative weight of CTB and UHB should vary between the Cu and O sites. However, our atomically resolved $dI/dV$ spectroscopy shows that the entire electronic structure is highly uniform in space without significant difference between the Cu, O, and Ca sites (Fig. S3). This suggests that both the CTB and UHB of parent cuprate form extended states in space rather than keeping their local orbital characteristics.

Now that the electronic structure of pristine CCOC is well established, we are ready to investigate the electronic state created by a single added charge. For this purpose we take advantage of the missing Cl defect naturally occurred on cleaved CCOC surface. Since each $\overline{Cl}$ ion strongly grasps an extra electron, a missing Cl effectively donates an electron into the $CuO_2$ plane. The red curve in Fig. 2c shows the *dI/dV* measured directly on the missing Cl site compared to that taken at a location far from the defect (black curve). The most dramatic spectroscopic feature caused by the doped electron is the emergence of a broad electronic state within the charge transfer gap near the edge of the UHB. Meanwhile, the peak of the UHB is completely suppressed, indicating the transfer of its spectral weight to the in-gap state. The CTB, in contrast, only shifts very slightly, indicating a strong particle-hole asymmetry induced by the doped electron. Moreover, the local charge transfer gap at the defect center is substantially reduced to $\Delta = 1.6 eV$. The schematic electronic structure induced by the doped electron is drawn in Fig. 2d.

The spatial distribution of the doped electronic state is another fundamental issue that STM is particularly suited for investigating. Figure 3a shows a series of *dI/dV* spectra taken as the STM tip is positioned away from the defect center along the Cu-O bond direction (inset of Fig. 3a), which reveals the vanishing in-gap state accompanied by the emerging UHB peak. At a distance of six lattice constants from the defect center, the *dI/dV* curve fully recovers that of the pristine Mott insulator. By subtracting this background from the *dI/dV* curve at each location, the relative contribution of the defect state can be extracted. As shown in Fig. 3b, both the amplitude of the in-gap state and the suppression of UHB decrease systematically with increasing distance from the defect center. Figure 3c shows that the profile can be fit very well by using two Gaussian terms representing the created in-gap state and suppressed

UHB respectively (see Supplementary Information for details). The spectral weight of the in-gap state can be roughly estimated by the area enclosed by the Gaussian fit (Fig. S4). Figure 3d summarizes the spatial evolution of the spectral weight, which loses half of its strength within a distance of three lattice sites, or merely 12Å from the defect center. The in-gap state is thus strongly localized in space.

High resolution topography near the defect reveals another interesting trend. Figure 4a shows the image taken at negative bias $V_s$ = -3.0V, which reflects the occupied state characteristics. Except for the dark pit at the missing Cl site, there is no other pronounced spatial pattern. Figure 4b displays the image taken at the same area with positive bias $V_s$ = 3.5V, which reflects the empty state characteristics. It shows a very dramatic spatial pattern where the four next-nearest-neighbor (NNN) sites appear to be anomalously bright. The strong bias dependence provides another piece of evidence for the particle-hole asymmetry where the CTB is insensitive to the doped electron but the UHB is strongly affected. Interestingly, similar features have been found at Ni impurities in superconducting $Bi_2Sr_2CaCu_2O_{8+\delta}$, where the NNN sites brighten up in the positive bias density of state map[11]. However, the clear contrast reversal at negative bias observed there is absent in the present case.

The properties of parent and lightly doped cuprates are crucial to the mechanism of high $T_C$ and have attracted tremendous theoretical interests[1]. It was first proposed by Anderson that the essential physics is contained in the half-filled Hubbard model[2]:

$$H = -t \sum_{\langle i,j \rangle \sigma} c_{i\sigma}^+ c_{j\sigma} + U \sum_i n_{i\uparrow} n_{i\downarrow} .$$

Here $t$ is the hopping integral between nearest-neighbor (NN) sites and $U$ is the onsite Coulomb repulsion. In the large $U$ limit as in the cuprates, the Hubbard model is transformed into the $t$-$J$ model, where $J = 4t^2/U$ is the superexchange energy. Since the two dimensional Hubbard or $t$-$J$ models have yet to be solved exactly, various approximation schemes or numerical methods are employed to extract the effective low energy physics. The STM results shown above provide unprecedented real space information and full electronic spectrum that could help discriminate the different models.

A major theoretical debate is whether a single band model is a valid low energy model for the cuprates[22]. The charge transfer nature of the parent insulator suggests that a three band model involving two O bands and one Cu band might be more appropriate. However, it was proposed that the lowest hole-doped state, namely the CTB, can be represented by a single Zhang-Rice singlet (ZRS) band[23], in which the doped hole residing resonantly in the four planar O sites form a singlet with the central $Cu^{2+}$ ion. There are two specific predictions in the ZRS picture. First, the Mott insulator phase has particle-hole symmetry and second, the ZRS can move freely through the $Cu^{2+}$ lattice and thus should distribute uniformly in space. Our STM results show that indeed undoped CCOC has a nearly particle-hole symmetric and spatially uniform electronic structure, apparently supporting the single band ZRS model. More quantitatively, the bandwidth of the CTB and UHB estimated from the Gaussian fit is in the order of $3J$ ($J = 0.13$ eV in cuprate), which is also close to the calculation based on the single band $t$-$J$ model[24].

The electronic state induced by a doped charge has also been investigated extensively by theory[25], but it is still under debate whether the main effect is the shift of $E_F$ or the emergence of new states or both. Our STM results demonstrate that the doped electron induces an in-gap state near the upper gap edge with its spectral weight mainly transferred from the UHB. There is no significant change of the Fermi level position and the CTB has negligible contribution to the spectral weight transfer. This rather unusual scenario is unexpected for doped charge transfer insulators[26,27]. We note that the electron donated by a surface defect as studied here may not be the ideal case of electron doping discussed by most theories due to the possible redistribution of charge on the surface, the change of local bonding situation, and the existence of a defect potential. Nevertheless, the local electronic structure of a surface electron-donor defect represents a well-defined problem that could be used for testing the validity of available theories for doped Mott insulators.

The peculiar spatial pattern near the defect reveals that the distribution of the doped electronic state is intimately related to the AF spin background. Figure 4c displays the schematic spin structure of the $CuO_2$ plane, showing that the NN and NNN sites constitute two magnetic sublattices. Figure 4d shows that hopping of the added electron to NN site will induce parallel alignment of NN spins which is energetically costly. Electron hopping to the NNN sites, on the other hand, preserves the AF spin configuration and is energetically more favorable (Fig. 4e). The brightness of the NNN sites in the positive bias image may reflect the significance of NNN hopping of the added electron. Although this simple picture is conceptually appealing, we reemphasize that a realistic model for the spatial structures near the defect, including the bias-dependent images and localization of the in-gap state, should include the influence of the defect in a similar manner as the treatment of impurity-induced state in superconducting cuprates[28].

The overall STM results support the single band $t$-$J$ model supplemented by the higher order hopping terms, which has been shown to capture the essential features of the ARPES and RIXS results[5,29]. More importantly, the high-bias scanning tunneling spectroscopy covering both the filled and empty states represents a powerful method for mapping the local electronic evolution. Applying this technique to cuprates with increased bulk dopings may help elucidate how the Mottness physics eventually leads to high $T_C$ superconductivity.

## Method Summary

**Sample growth.** The single crystals of $Ca_2CuO_2Cl_2$ (CCOC) are grown by the flux method as described elsewhere[30]. The powders of CaO and $CuCl_2$ are mixed with a molar ratio of 2:1 and put into an alumina crucible. The mixed powder is heated at 1053K for 24 hours (hs) with intermediate grindings. After that, the CCOC precursor is heated to 1053K at a ramp rate of 60K/h and kept at this temperature for 5h. It is then heated to 1203K at a ramp rate of 60K/h and kept at that temperature for 10h. In the end it is cooled down to room temperature at a ramp rate of 60K/h. Single crystals with typical size of 2mm×2mm×0.1mm can be harvested by cleaving the as-grown bulks.

**STM measurements.** The STM experiments are performed with a cryogenic variable temperature ultrahigh vacuum STM system. The CCOC crystal is cleaved *in situ* at $T = 77$K and then transferred immediately into the STM sample stage. To make tunneling into the insulating CCOC possible, the STM experiments have to be performed at relative high temperature so that the sample has finite conductivity. At $T = 5$K the STM tip will crash into the sample because the sample is so insulating that no tunneling current can be detected. Therefore all STM results reported in this paper are acquired at $T = 77$K. In addition to that, the sample bias voltage has to be set beyond the charge transfer gap so that there is finite density of state in the sample. Moreover, a small tunneling current is used because otherwise the tip-sample junction will be quite unstable under large biases at $T = 77$K. The typical setup parameters are sample bias $V_s = \pm 3$V and tunneling current $I_t = 2\text{-}10$pA. The STM topography is taken in the constant current mode, and the $dI/dV$ spectra on CCOC are collected using a standard lock-in technique with modulation frequency $f = 423$Hz.

**Supplementary information:** A supplementary information session is available.

**Acknowledgments** We thank Z. Y. Weng, T. Xiang, F. C. Zhang, and G. M. Zhang for helpful discussions. This work is supported by the NSF of China and by the MOST of China (Grants No. 2009CB929402 and No. 2010CB923003).

**Author Information** Correspondence and requests for materials should be addressed to Y.W. (yayuwang@tsinghua.edu.cn).

Figure Captions:

**Figure 1 | Crystal structure and STM topography on CCOC**. **a**, Crystal structure of $Ca_2CuO_2Cl_2$, which is similar to that of $La_2CuO_4$ except that the LaO layer is replaced by CaCl layer. The crystal cleaves easily between two adjacent CaCl layers, as indicated by the gray planes, exposing the square lattice of Cl atoms. **b**, Constant current topographic image of CCOC taken at bias voltage $V_s$ = -3.5V and tunneling current $I_t$ = 10pA. Each cross-shaped dark site on the surface corresponds to a missing Cl defect most likely induced by the cleaving process.

**Figure 2 | Spectroscopy on pure CCOC and the missing Cl defect**. **a**, Spatially averaged *dI/dV* spectrum over a defect-free area of CCOC reveals the full electronic structure across the Mott gap. (Inset) High-resolution image of the area where the spectra are taken. **b,** Schematic band structure of the pristine Mott insulator showing the upper Hubbard band (unfilled), the charge transfer band (pink) and the non bonding O $2p_\pi$ band (green). **c**, The red curve shows the *dI/dV* spectrum measured at the missing Cl defect center and the black curve is taken at a location far from the defect. (Inset) High-resolution image around a missing Cl defect and the red dot indicates where the red curve is taken. **d**, Schematic electronic structure at the defect center: an in-gap state (blue) emerges near the edge of UHB due to spectral weight transfer from the latter. The CTB is slightly shifted and the local gap amplitude is significantly reduced.

**Figure 3 | Spatial distribution of the doped electronic state. a**, A series of *dI/dV* spectra taken at locations away from the defect center along the Cu-O bond direction, as indicated in the inset by colored dots. **b**, Contribution of the doped electronic state to local electronic structure obtained by subtracting the background *dI/dV* (obtained at six lattice distances away from the defect) from each *dI/dV* spectrum in **a**. **c**, The *dI/dV* curves in **b** can be fit well by two Gaussian terms and the spectral weight of the in-gap state can be estimated by the area enclosed by the Gaussian lineshape. **d**, Spatial dependence of the in-gap state spectral weight,

which decays rapidly with increasing lateral distance from the defect center.

**Figure 4 | Bias dependent images around the defect. a** and **b**, Constant current images around a missing Cl defect taken at negative and positive biases respectively. The positive bias image shows four bright spots at the NNN sites of the defect. **c** to **e**, Schematic illustration of the spin structure of the Mott insulator and how an added electron hops in the presence of an AF spin background. Hopping to the NN site ($t$) disturbs the original AF spin configuration (the red cross indicates the parallel alignment of NN spins), whereas hopping to the NNN sites ($t'$) has no spin consequence.

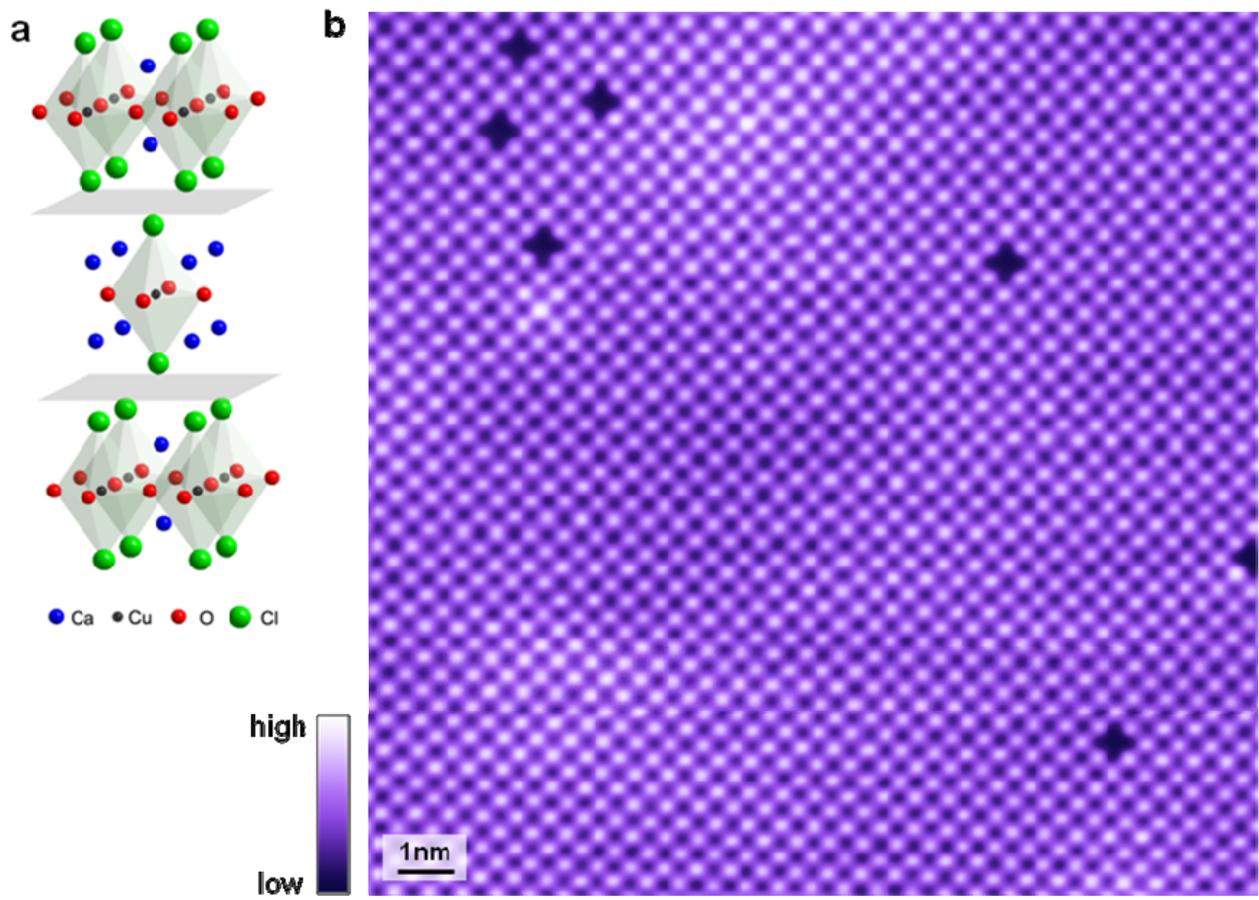

Figure 1

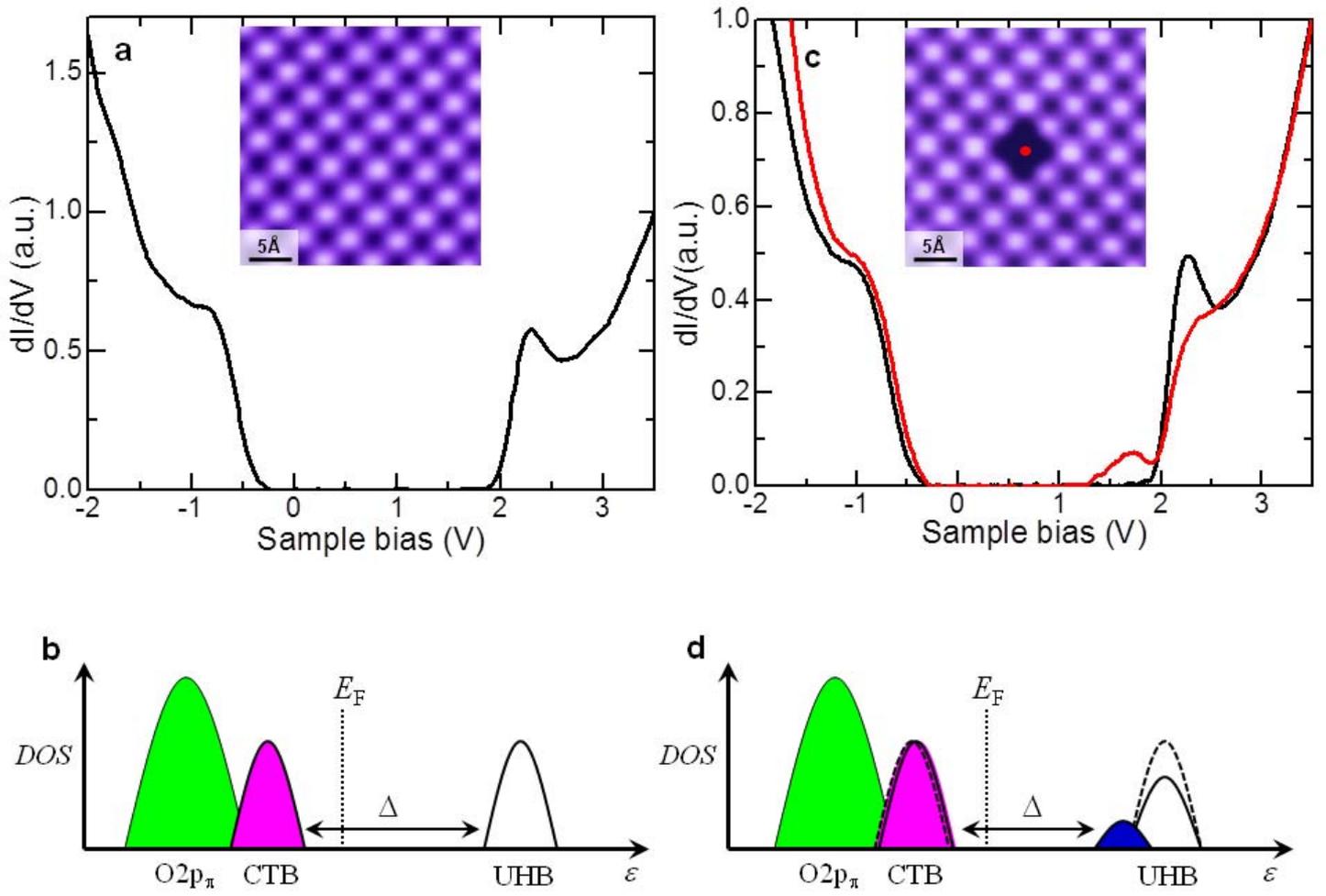

Figure 2

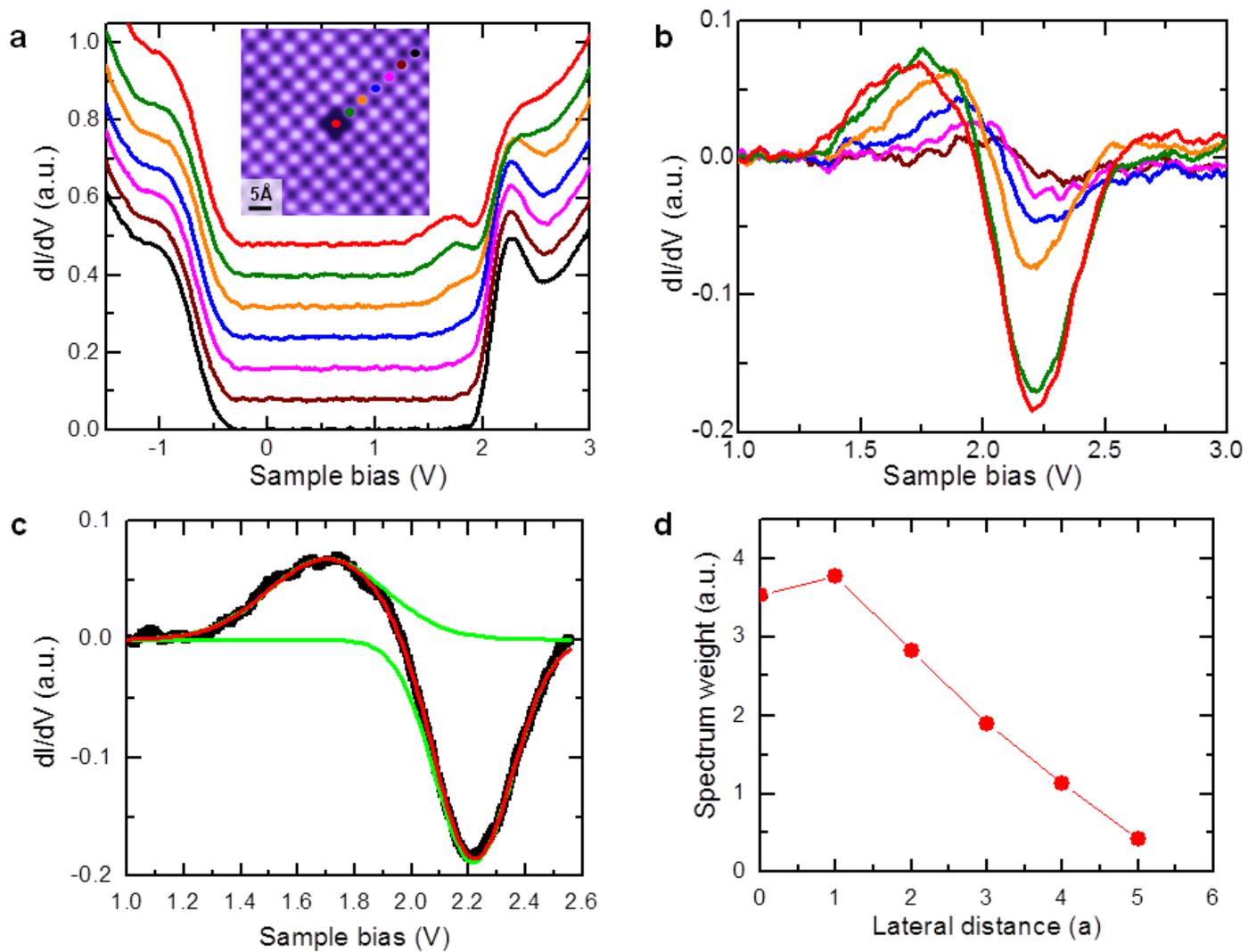

Figure 3

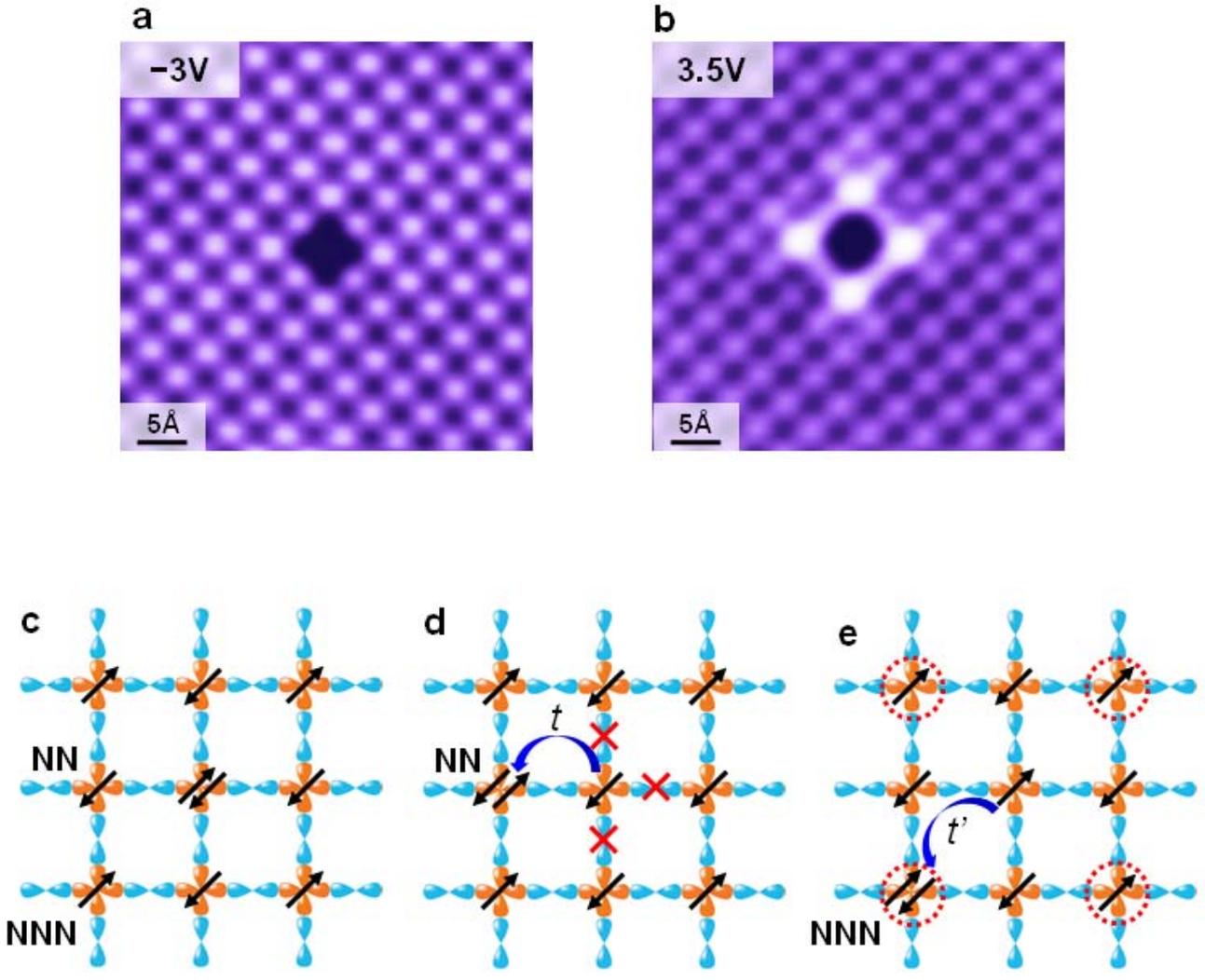

Figure 4

Supplementary information for

# Visualizing the atomic scale electronic structure of the $Ca_2CuO_2Cl_2$ Mott insulator

Cun Ye[1,*], Peng Cai[1,*], Runze Yu[2], Xiaodong Zhou[1], Wei Ruan[1], Qingqing Liu[2], Changqing Jin[2] & Yayu Wang[1,†]

[1]*State Key Laboratory of Low Dimensional Quantum Physics, Department of Physics, Tsinghua University, Beijing 100084, China*

[2] *Institute of Physics, Chinese Academy of Sciences, Beijing 100190, China*

[*]*These authors contributed equally to this work.*

[†] E-mail: yayuwang@tsinghua.edu.cn

## Contents:

**SIA: Preparation and calibration of the STM tip**

**SIB: Gaussian fit to the *dI/dV* spectra of the CTB and UHB**

**SIC: Atomically resolved electronic structure on pure CCOC**

**SID: Fitting of the in-gap state spectra**

**Figure S1 to S4**

**References**

# SIA: Preparation and calibration of the STM tip

The STM tip condition is of particular concern when addressing the high-energy spectroscopic features. Not only has it to be stable under high biases, but also its electronic DOS should be relatively featureless over a wide energy window. We have developed a routine for STM tip treatment and calibration. We use electro-chemically etched polycrystalline tungsten tip for all the measurement. Before the STM experiments on CCOC, the tip is first treated by *in situ* high voltage e-beam sputtering, which is very efficient for cleaning and sharpening the tip. We then use the tip to approach an atomically clean crystalline Au(111) surface prepared *in situ* by repeated Ne ion sputtering and high temperature annealing. By taking topographic scans and spectroscopic measurements on clean Au(111) surface, we could check if the tip shape or spectrum is ideal. If necessary further tip treatment can be done by field emission process or controlled gentle crash into the Au(111) crystal.

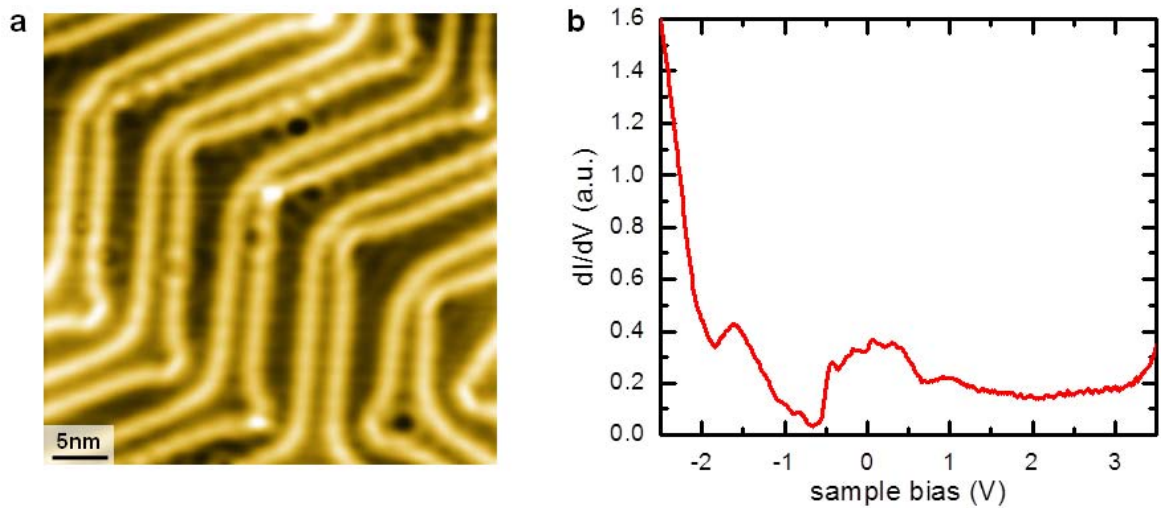

Fig.S1. **a**, Constant-current STM image of Au(111) surface ($V_s$ = 0.1V, $I_t$ = 0.1nA), which shows clearly the herringbone reconstruction. The standing wave patterns near the defect are also visible. **b**, Typical *dI/dV* spectrum measured with an clean tungsten tip on Au(111) surface, which shows no distinctive feature except the surface state band setting in around $V_s$ = −0.5V

Fig. S1a shows a typical constant-current image on Au(111) surface taken with $V_s$ = 0.1V

and $I_t$ = 0.1nA, which shows clear herringbone reconstruction, standing waves patterns induced by the quasiparticle interference of the surface state, and nearly circular shaped defects. Fig. S1b displays the *dI/dV* spectroscopy over a wide range of sample bias from -2.5V to 3.5V. The abrupt jump at $V_s$ ~ −0.5V is due to the onset of the surface state band of the Au(111) surface[1]. The steep increase of *dI/dV* at large negative bias is due to the Au bulk electronic states. There is no spurious electronic state within the energy range of interest that may be induced by unwanted adsorbates on the tip. Both the topography and spectroscopy indicate that the tip is clean and stable. High quality large-bias, high temperature STM results can usually be obtained with such a tungsten tip.

## SIB: Gaussian fit to the *dI/dV* spectra of the CTB and UHB

In previous ARPES studies on CCOC, it was found that the energy distribution curve (EDC) along a specific momentum (*k*) of the charge transfer (CTB) or Zhang-Rice singlet (ZRS) band can be fit by a broad Gaussian lineshape rather than a sharp Lorentzian lineshape for well-defined quasiparticles (qps). Moreover, the width of the EDC is comparable to the entire bandwidth determined by the dispersion of the band[2]. These observations have been interpreted as evidence for the breakdown of the coherent qp picture in the Mott insulators.

Unlike the *k*-resolved EDC curve measured by ARPES, the *dI/dV* curves measured by STM are related to the *k*-integrated spectral function. Interestingly, we find that the *dI/dV* curves of the CTB and the upper Hubbard band (UHB) on CCOC shown in the main text can also be fit by Gaussian lineshape. Fig. S2a shows the zoom-in *dI/dV* curve of the CTB, which shows that the low energy half of the curve can be fit very well with a Gaussian lineshape. For the high energy half, the fitting is spoiled by the rapid increase of *dI/dV* due to the onset of the nonbonding oxygen states. The width of CTB, estimated to be 0.47eV from the full width at half maximum (FWHM) of the Gaussian fit, is in rough agreement with the bandwidth (*W* = 0.35eV) determined by ARPES. The situation of the UHB, which cannot be detected by ARPES, is quite similar. Fig. S2b shows the *dI/dV* curve of the UHB, which can also be fit very well by the Gaussian lineshape with a FWHM of 0.34eV. Both the band width of the

CTB and UHB thus is in the order of 3$J$ ($J$ = 0.13eV in the cuprates), the calculated band width of the parent cuprate based on the single band $t$-$J$ model[3].

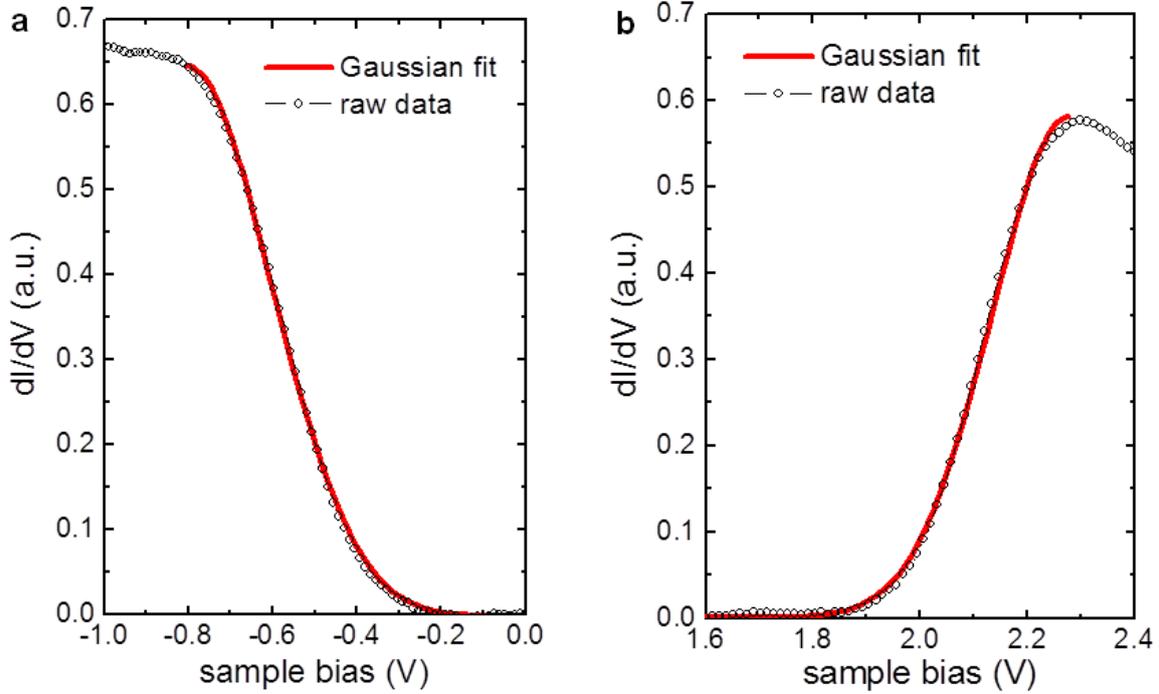

Fig.S2. The *dI/dV* spectroscopy of the CTB (a) and UHB (b) of undoped CCOC. The red curves indicate Gaussian fits to the data.

## SIC: Atomically resolved electronic structure on pure CCOC

The real space electronic characteristics are of great importance to Mott insulators, where local Coulomb repulsion plays a dominant role. STM is arguably the most ideal experimental technique for addressing this issue. Fig. S3 displays spatially resolved *dI/dV* spectra taken at three different atomic sites, as marked in the inset. The black, red and blue curves depict the spectra taken on Cu, O and Ca sites, respectively. The setup conditions are $V_S$ = −3V and $I_t$ = 10pA. The curves are barely distinguishable on the negative bias side. In particular, the CTB possesses exactly the same peak shape and intensity on these 3 sites. On the positive bias side, the intensity of UHB is slightly suppressed on Ca site compared to those on Cu and O sites. However, the difference is rather small and varies slightly from different experimental runs.

Therefore we could conclude that there is no significant difference between the electronic structures on different lattice sites. Both the CTB and UHB are uniformly distributed in space.

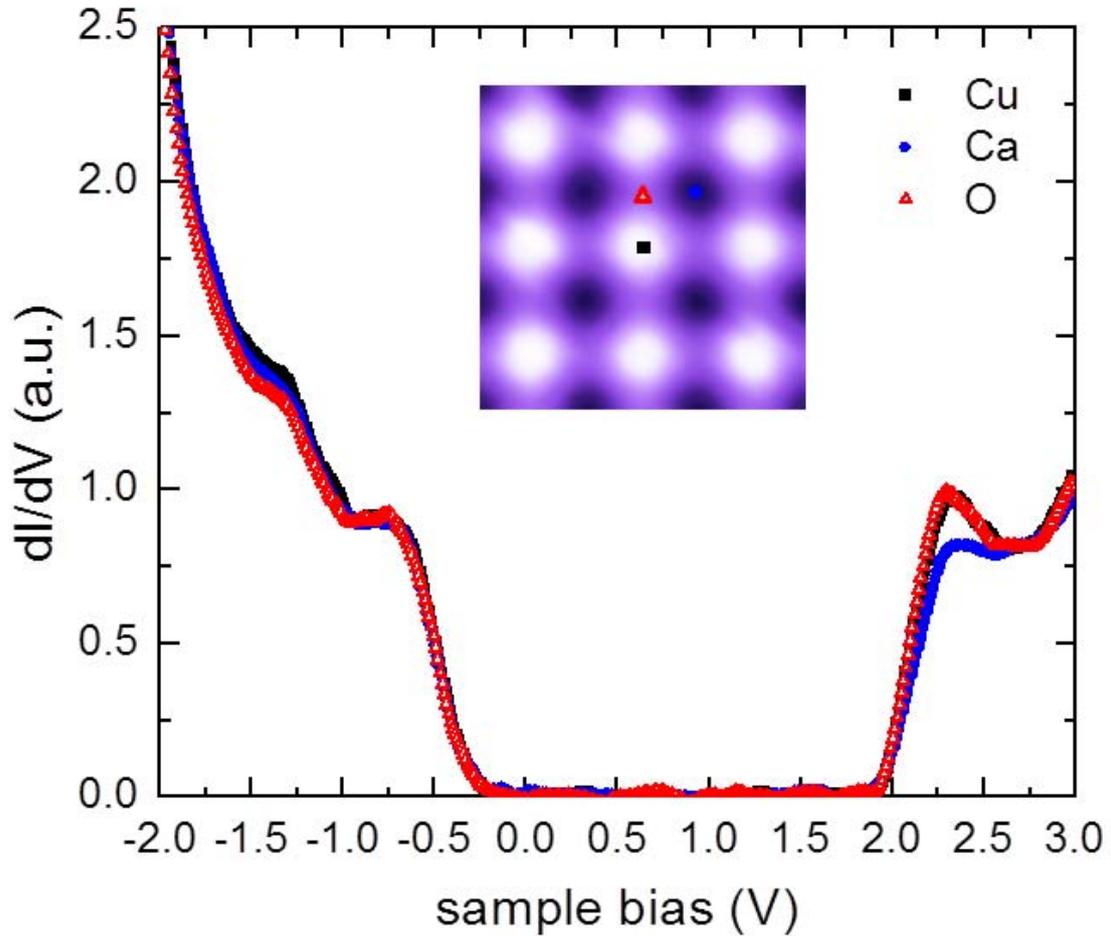

Fig.S3. *dI/dV* spectra taken on the Cu, Ca and O sites of the CCOC surface, as indicated in the inset. There is no significant difference between the spectra, demonstrating spatially uniform electronic structures.

## SID: Fitting of the in-gap state spectra

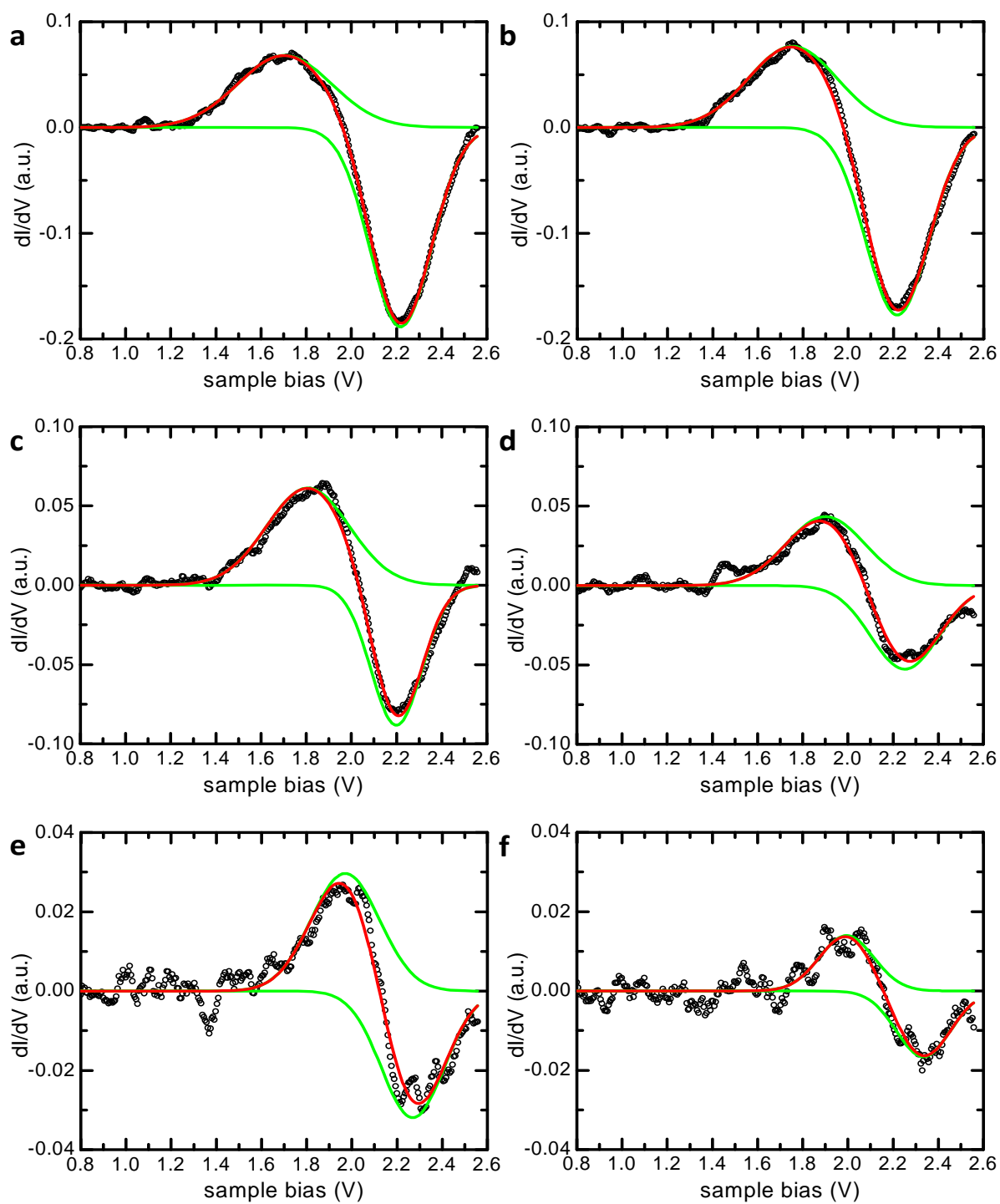

Fig.S4. (**a** to **f**) *dI/dV* spectra taken at locations away from the missing-Cl along the Cu-O bond direction with the background subtracted. The green curves show the two Gaussian curves for the in-gap state and the suppression of UHB respectively, and red curves show the summation of the two terms.

Fig. 3a in the main text displays the electronic structure measured at different locations near the missing Cl defect, which demonstrates the spatial decay. By subtracting *dI/dV* curve of pristine CCOC from spectrum measured around the defect (Fig. 3b), the relative contribution of the defect state and the suppression of UHB can be extracted. We find that the subtracted curves can be fit very well with the following formula:

$$y = \sum_{i=1,2} \frac{A_i}{\sqrt{\pi}\sigma_i} \exp\left(-\frac{(x-x_{ci})^2}{\sigma_i^2}\right)$$

Here the summation includes the two Gaussian terms representing the created in-gap state and the suppression of UHB. $|A_i|$ corresponds to the area enclosed by each curve, $x_{ci}$ is the peak position and $\sigma_i = \Gamma_i/\sqrt{4\ln 2}$ where $\Gamma_i$ is the FWHM of the Gaussian fit. The green solid curves in Fig. S4 (a to f) indicate the two Gaussian curves while the red curves show the combination of the two terms. The agreement between the fit and the data is very impressive. The in-gap state spectral weight plotted in Fig. 3d of the main text is the $|A_1|$ value, i.e., the area enclosed by the Gaussian fit for the created in-gap state.